# STAR-Scheduler: A Batch Job Scheduler for Distributed I/O Intensive Applications

V. Mandapaka[(a)], C. Pruneau[(b)], J. Lauret[(c)], S. Zeadally[(a)]

[(a)] Department of Computer Science, Wayne State University
[(b)] Department of Physics and Astronomy, Wayne State University
[(c)] Physics Department, Brookhaven National Laboratory

Abstract

We present the implementation of a batch job scheduler designed for single-point management of distributed tasks on a multi-node compute farm. The scheduler uses the notion of a meta-job to launch large computing tasks simultaneously on many nodes from a single user command. Job scheduling on specific computing nodes is predicated on the availability of user specified data files co-located with the CPUs where the analysis is meant to take place. Large I/O intensive data analyses may thus be efficiently conducted on multiple CPUs without the limitations implied by finite LAN or WAN bandwidths. Although this Scheduler was developed specifically for the STAR Collaboration at Brookhaven National Laboratory, its design is sufficiently general, it can be adapted to virtually any other data analysis tasks carried out by large scientific collaborations.

## 1. *Introduction*

Data analyses conducted by large physics experiments, such as the STAR [1] experiment at Brookhaven National Laboratory, require repeated processing of large Giga to Terabyte data sets. Typically, the same processing task must be repeated numerous times to analyze data units, called events, acquired experimentally, or generated through complex simulation codes. For STAR, the data reduction and analysis is operated in two phases. The first phase consists of the calibration and reconstruction of the data. It is accomplished at the RHIC Central Reconstruction Farm with dedicated analysis software devised by the collaboration. The reconstructed data are stored in data summary files in a central repository at the RHIC Computing Facility. These files thence form the basis of physicist/users analyses, and are processed repeatedly, as a second phase analysis to extract physics observables and results. Typically from a few tens to many hundreds of files are then processed: the same analysis code is run on multiple thousands, often millions, of events to generate histograms, and extract specific physics quantities. Event sizes may range from kilobytes to a few megabytes, and depending on analysis specificities, their processing may require from a few milliseconds to many tens of seconds per event. The problem is also compounded by the fact that many tens of users, typically, wish to process the same datasets and conduct different physics analyses at the same time. Thus, although specific analyses may not constitute large tasks, the need to share data files amongst many users, and the repeated access to the same files ends up creating an I/O bottleneck that can severely limit the efficiency and speed of user analyses if files are located on, and served by a single server.

Given the current relatively low price of CPUs and hard disks, we have opted to implement a system where many tens of CPUs are each equipped with multiple large disks. Phase two data files are then copied from the central archive to the distributed disks. Files in high demand are further copied at multiple nodes to enable multiple concurrent analyses. As jobs may then be partitioned and executed on CPUs collocated with the data to analyze, only a limited amount of network traffic is required during user analyses. I/O is optimized, and CPU efficiency is maximized.

The drawback of this approach lies however in that users need to find where their data files are located, and thence submit multiple equivalent analysis tasks on the identified nodes. Although the analyses are likely to be completed in principle more rapidly, the tasks of users become more onerous, if not tedious, and the overall performance improvement might then be considered negligible unless tools are available for users to easily locate datasets, and submit multiple jobs from a single control point. It was the purpose of this work to develop a job scheduler that permits query to a data file catalog to identify file locations, and transparent submission of a user selected analysis task involving one or many computing nodes, from a single command, using a simple meta-job description language.

An overview of the Scheduler design is presented in Section 2. It is followed by implementation specifics in Section 3. Finally, in Section 4, a summary and future course of research are presented. A more in depth description of the design, implementation, and technical aspects of the scheduler is presented in [2].

## 2. *Design Overview*

The primary goal of the STAR-Scheduler is to provide a user friendly interface for transparent job submission of analysis tasks to be distributed on one, or many computing nodes. The user should not need to know or care where the data to be analyzed is located, or what computing nodes will be used for data processing. The user shall be able to specify the data to analyze using generic keywords descriptive of the specific dataset to process, i.e. data from what year, what colliding system (e.g. Au + Au nuclei), the beam energy, the detector and trigger configurations, etc.

The scheduler is required to be flexible, adaptable, and portable on virtually any operating system architecture. We thus opted to proceed to its development in JAVA language, using an object-oriented design. JAVA is by design platform independent, and thereby makes ports to different architecture relatively simple. An object oriented philosophy enables for flexibility, growth, and easy maintenance.

A number of existing, publicly available JAVA standard interfaces, and packages, were used in the development and deployment of the scheduler. The Java Document Object Model (JDOM) public package [3], and APACHE SAX XERCES package [4] are used for parsing user input provided in the form of XML documents. Use of JDOM required the development of an XML dictionary for job submission language and protocol. The XML dictionary we developed is illustrated by an example of user input file in Appendix A.

The Scheduler was designed to enable user data file selection from either a database query or from a user defined file list. The STAR deployment of this scheduler was extended to utilize the STAR File Catalog and query system [5]. In this work however, we focused on the development of the scheduler prototype and therefore limited user inputs to file lists.

The design of the STAR-Scheduler was realized with the help of Unified Modeling Language (UML) tools. Multiple Use-case scenarios were generated in the design stage to analyze the various operations and states of the Scheduler. The use-case scenarios were then used to identify the specific components of the scheduler, and subsequently proceed to class designs. The Scheduler follows a hierarchical design pattern for class diagrams. Interfaces were first declared to define the scheduler actions, and behavior. Concrete classes were then implemented in Java as sub-classes of the interfaces to realize the behavior represented by these abstract interfaces.

A schematic overview of the Scheduler operation is shown in Fig 1. The goal is for the user to be able to submit a complex job request, possibly involving execution of a program on many distinct compute nodes, using a single “meta” command. The user should not have to micro-manage the details of which or how many compute node(s) will actually be involved in the process. Submission of a job should be possible based on a program identifier, and an abstract specifier of the data to be analyzed. The task of the scheduler is thus to analyze the user request, identify what program to execute, what data files are to be analyzed, and spawn jobs on the nodes where the data are located to execute the program selected by the user to analyze the data found on those nodes.

The Scheduler operation is best exemplified by the following use-case. A user submits a job request via a call to a scheduler peer. The submission may, in principle, be accomplished by means of different sources and mechanisms such a database, flat text file, XML file, or the output from a different program. In this work, the implementation of the Scheduler was restricted to consider only XML files as input. Alternative input formats may however be trivially accomplished with proper extension of the scheduler input module. The Scheduler receives the request from the client and puts it in a queue for sequential processing. The scheduler analyzes and validates the received requests according to internal syntactic and semantic rules. Processing of a request is aborted if a request is found invalid, and the user is so notified. If valid, the scheduler proceeds to generate a meta-job request. A meta-job request, heretofore called MJR, encapsulates the actions to be taken by the Scheduler to fulfill the user's request. It describes the program to execute, parameters to be passed to this program, data to process, conditioning attributes, etc. To generate an MJR, the Scheduler first identifies the node location of the different data files to be processed. Each file to be processes is represented by a file element, which contains its name, location, and various other attributes. File elements are sequentially added to the MJR to achieve a complete description of the task to accomplish. Once completed, the JobRequestAnalyzer analyzes the MJR. It uses a policy dictated by the system administrator, to build *ElementaryJobRequests* (EJR) encapsulating specific jobs to be eventually dispatched on remote computing nodes. The generated EJR each contain a subset of file elements present in MJR. The policy defines rules for the creation of new elementary job requests with node property determined according to the environment of the cluster.

The analyzed MJR is submitted by the Scheduler to the Meta-Job-Dispatcher. The Meta-Job-Dispatcher establishes communication with the remote nodes and submits EJRs for execution. In the event of successful submission of job requests to remote nodes, the dispatcher proceeds to the next request. If a failure occurs, an error log is created for the failed job request. In this work, a Java-RMI engine is used to establish communication with remote nodes and adding the job request to their local queue. It is however possible to use a different mechanism for remote job submission. For instance, the Scheduler may be configured to dispatch job requests to queue management systems such as LSF [6], PBS [ 7], or NQS [ 9], using appropriate syntaxes, thereby reducing the burden on the STAR-Scheduler for queue management and optimization.

The EJRs are queued on the "remote" nodes for execution. Dispatchers running on remote nodes query their local queue manager for the next job request. The queue managers hold references to all the queues running on their local node. They assign jobs for execution on appropriate queues. The dispatchers execute the program declared in the job request, with user-selected parameters, and data files identified by the scheduler. Upon successful completion, the program's output is saved at a location specified by the user. Instead, if a failure occurs, the program is aborted, and an error log is produced.

The scheduler is designed so the above operations are transparent to the user submitting job requests. A user submits a job request to the Scheduler as though he/she is executing the program on the local node where he/she is logged on. Other functions are also available. It is for instance possible to query the scheduler to determine the current state of a job, i.e. whether the job is pending, submitted, queued, completed or possibly aborted. The "pending" state corresponds to the job request being held on the local host for submission to remote nodes. The "submitted" state corresponds to the job request being submitted to remote nodes for execution. A job request queued for execution on remote node is identified as "queued". If a job is successfully completed, it is shown as "completed" or else as "aborted

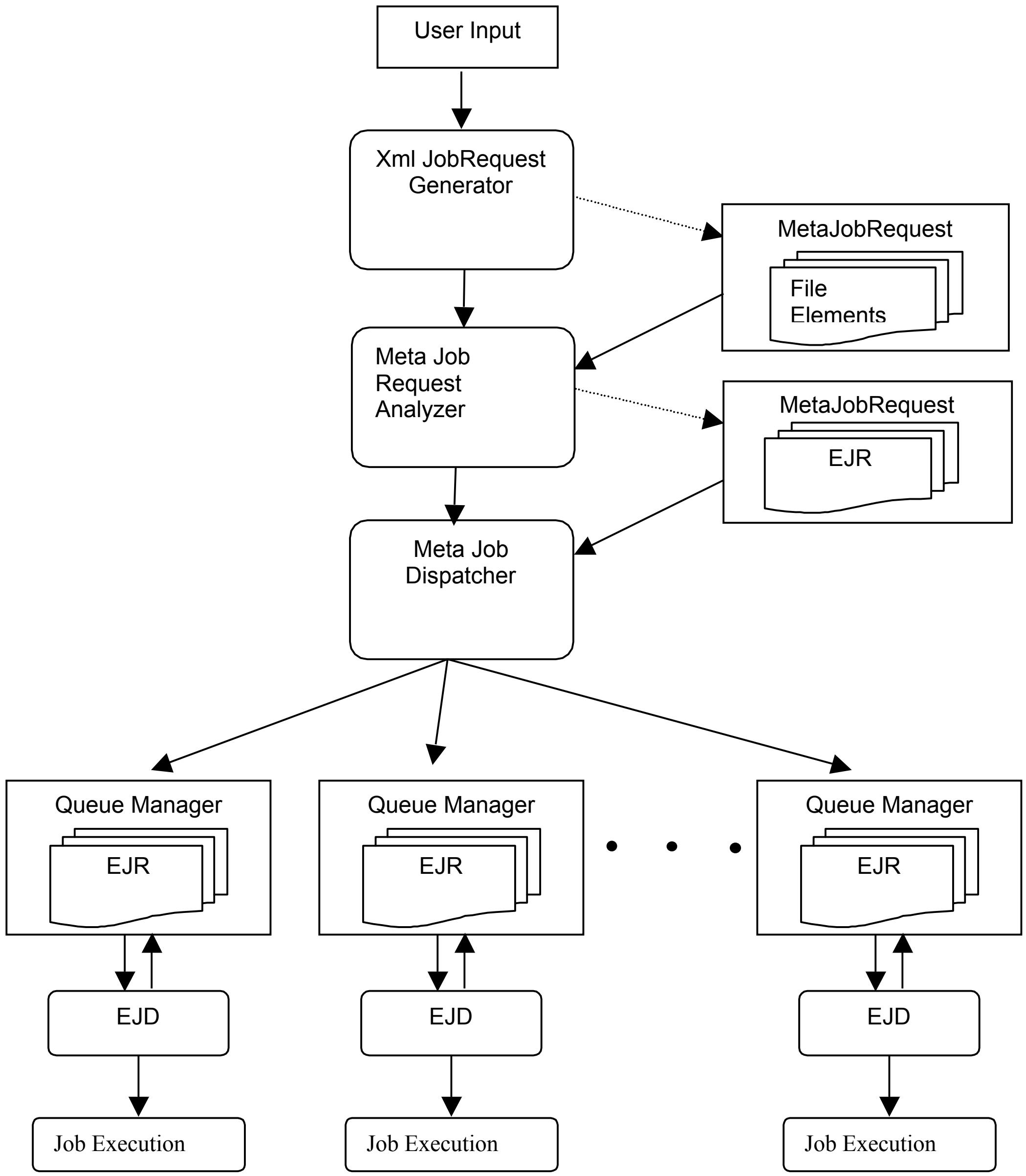

Figure 1: Overview of the Scheduler Design. EJRs are Elementary Job Requests; MJR are Meta-Job Requests. EJD stands for Elementary Job Dispatcher. See text for details.

## 3. *Implementation of STAR-Scheduler*

The Scheduler implementation consists of one main package and six sub-packages. Their names, relationships, and dependencies are illustrated in Fig 2. The top-level package, called *scheduler*, defines and includes the high-level classes, which implement the functionality of the Scheduler. The sub-packages *job*, *queue*, *system*, *comm*, *conf*, *dispatcher*, and *load* instrument specific aspects of the various actions, and tasks related to job scheduling. Sub-packages include an abstract interface defining the package functionality, and a number of concrete classes implementing this functionality. The different packages, interfaces, and implementations are briefly summarized in the following.

The top-level package *scheduler* is primarily based on the *JobRequestGenerator* interface which defines methods for the generation of MJR. The generation of MJR in response to a user request is defined in the implementing classes specific to the type of user input.

The *JobRequestGenerator* defines the method *constructJobRequest*, which encapsulates the notion of creating an instance of an MJR. Subclasses extending this interface implement this method for a specific input format to achieve the task of creating an MJR instance. For XML input type, the class *XmlJobReqGenerator* is used for the analysis of the user input and the generation of an MJR instance. The *XmlJobReqGenerator* implementation of the *constructJobRequest* first instantiates an MJR, and then proceed to parse the user input using standard parsing tools from the APACHE Xerces package [4], and sequentially add references to new file element objects. A sample user input is presented in Appendix A.

The notion of a user request is encapsulated in the JobRequest interface included in the package *job*. This interface forms the basis for notions of meta-job-requests describing a full analysis job to be submitted to a computing farm, and elementary-job-requests describing job executed on a specific computing node. The classes *MetaJobRequest* and *ElementaryJobRequest*, included in the *job* package, are both based on the *JobRequest* interface, and provide implementations to handle meta and elementary job requests respectively. The package *job* also includes the *JobRequestAnalyzer* interface that defines methods for the generation of EJRs on the basis of user inputs. The *MetaJobRequestAnalyzer* provides the implementation for the analyzer interface using the rules defined in the job request generation policy.

The *JobRequest* interface defines variables such as the program name to be executed, the name of the user submitting the job request, and the working directory of the executable program. It also defines methods for accessing and setting the above variables. The *MetaJobRequest* class inheriting the *JobRequest* interface provides the implementation for the above methods. In addition, the MJR also defines methods that handle references to multiple

*FileElement* and *ElementaryJobRequest* objects. The class *ElementaryJobRequest* implements the *JobRequest* interface to handle job submission to specific computing nodes. It carries additional data members to specify the host where the data is localized, and the job will be executed. It also includes a reference to the parent host issuing the request. Since the EJR is involved in communication over network, the class implements the java.io.Serializable interface [8] to support Java object serialization and deserialization features.

The JobRequestAnalyzer interface defines the *analyze* method to encapsulate the task of JobRequest analysis. The MetaJobRequestAnalyzer implements this method to provide for the analysis of MetaJobRequest using the Policy interface. The analysis proceeds by sequentially sorting the file elements according to their host name, and generating EJR instances for each new host encountered. References to the files located on the given hosts are then added to the corresponding EJRs. Subsequently the EJR themselves are added to the MJR.

The *JobDispatcher* interface features one method, *dispatch*, to encapsulate the notion of dispatching a *JobRequest*. The classes *MetaJobDispatcher* and *ElementaryJobDispatcher* provide implementations of this method to dispatch meta- and elementary job respectively. *MetaJobDispatcher* uses classes and tools of the *comm* package to establish inter-process communication and to queue EJRs on remote nodes. The *ElementaryJobDisaptcher* (EJD) class implements the *dispatch* method as a process running in its own thread. The thread awakes at regular 60 seconds intervals, and queries the *QueueManager*, defined in the *queue* package, for EJR awaiting execution, and submits the execution request to the runtime environment. The thread returns to sleep if there are no pending jobs in the queue. On successful job completion, the EJD dispatcher proceeds to the next EJR in queue for execution. In the event of a failure, a job error log is created.

Given that the Scheduler is developed in Java, it can in principle be used on any OS architecture. In practice, we have implemented the Scheduler to be able to use an existing batch queue system or its own job launcher and manager. Different institutions may use various batch queue systems such as PBS [7], LSF [6], NQS [9]. The code was therefore specifically designed to enable selection at runtime of the proper job queuing system. This is achieved by using an abstract interface to define the required behavior and interface between the Scheduler and various batch queue systems. The abstract interface is part of the *dispatcher* package. Specific implementations of this interface, appropriate for operation at BNL and Wayne State University, were deployed.

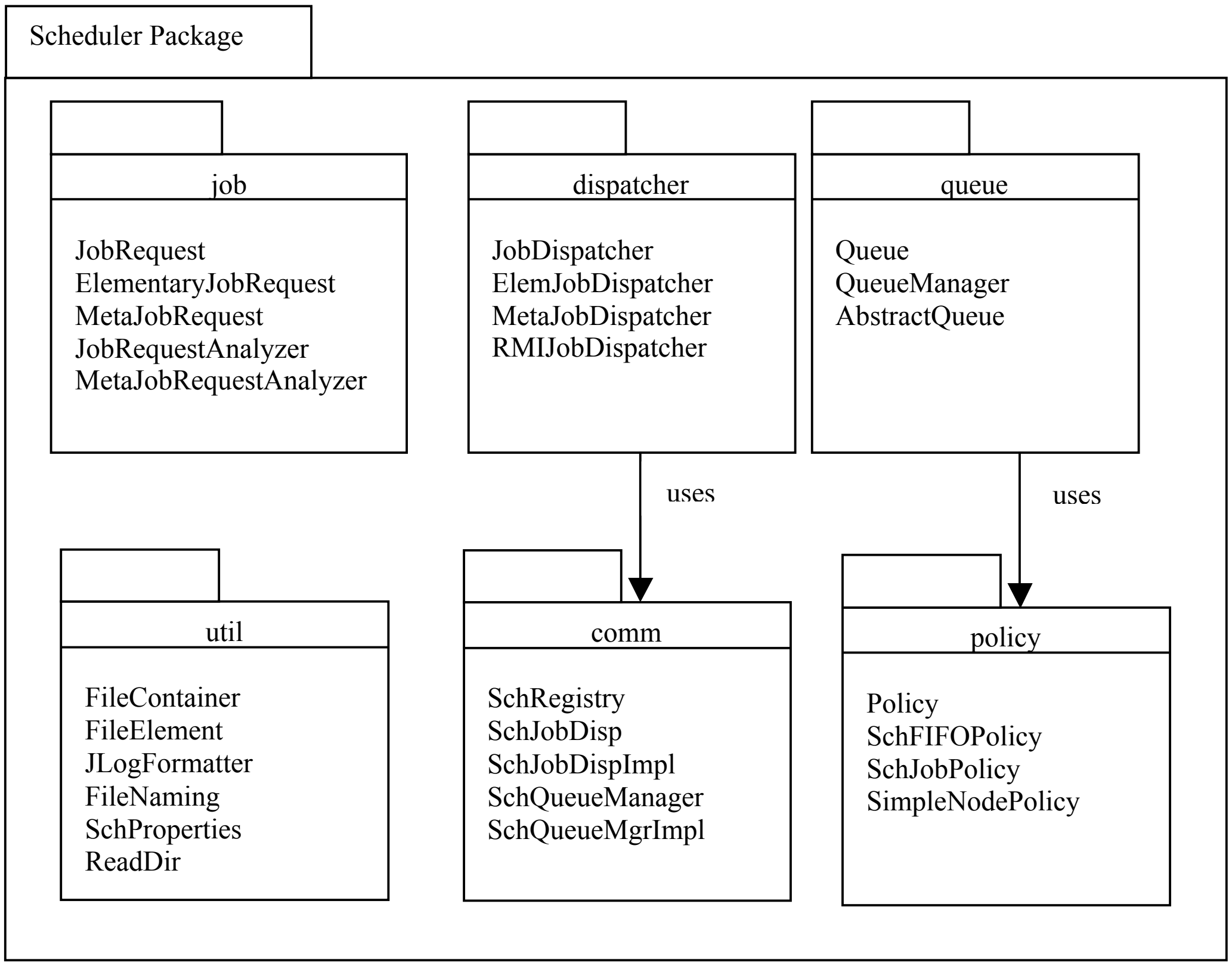

Figure 2 Scheduler package and classes overview.

## 4. *Summary*

We developed and deployed a first generation job scheduler for the submission, and management of high energy physics analysis jobs on a medium-size computing cluster operated on a local area network. We developed the notion of meta-job request, and an associated job submission language protocol in XML to enable submission of complex jobs, partitioned to run in parallel on tens of computing nodes, from a single user command. The scheduler basically acts as a top-level manager to determine where data to be analyzed are located, and to create, and dispatch jobs to be executed specifically on those computing nodes. This enables users to submit jobs simultaneously on multiple computing nodes, based on a single command, to analysis Giga or event Terabytes size data sets.

The scheduler was created in JAVA, and can as such be used on virtually any OS architecture. The software was initially written and tested on a small cluster of 10 LINUX machines connected via a 100 MB/s Ethernet network. It has now been extended and successfully deployed at Brookhaven National Laboratory by the STAR collaboration where it is used by a large group of physicists (>50) for analysis of Terabyte size datasets [5]. The test implementation at Wayne State University used a custom-made job queuing system to manage job submissions, whereas the STAR implementation uses the commercial LSF batch queue system. Other job queuing systems may also easily be integrated.

Users note their task is greatly simplified by the scheduler given large job submissions, job tracking, and monitoring of jobs involving tens of processors, can be accomplished trivially from a single point.

Possible extensions of the Scheduler include ports to WAN domains, and integration to GRID technologies. Of particular interest is the development of smart algorithms to automate the transfer and fair distribution of data within the fabric to further optimize the overall system performance.

## *APPENDIX A*

The following table presents an example of user input submitted to the scheduler to perform a data analysis. The XML input consists of a list of four parameters groups and their associated attributes.

The parameter *process* specifies the user and program to be executed. The tag *username* is used to specify the user name, while the *command, title* and *description* tags specify the name of the program or command to execute, the name and a short description of the job submission.

The parameter *stdin* is used to determine program inputs. The *name* attribute identifies command parameters to be passed on the command line while submitting the job to the operating system; the *path* attribute refers to the work directory where the program should be executed; and *machine* indicates the name of the computing node where these parameters apply.

The parameter *stdout* is used to determine program outputs. The *name* attribute is used to specify the file name where program output should be saved; the *path* attribute identifies the directory location of that file; and *machine* indicates the name of the computing node where these parameters apply.

The parameter *input* is used to specify a file that contains the complete list of data files to be analyzed by the job. The *name* and *path* attributes specify respectively the file names and directory location.

```xml
<? Xml version="1.0" encoding="UTF-8"?>
<process username="einstein" command="program_name" title="my_program"
description="a_simple_sample">
<stdin name="-r " path="/home/users/einstein/"
machine="node1.physics.wayne.edu" />
<stdout name="output.log" path="/home/user/einstein/output"
machine="node1.physics.wayne.edu" />
<input name="my_data.list" path="/usr/local/prog_input"
machine="node0.physics.wayne.edu" />
</process>
```

Table 1 : Example of Scheduler Input using an XML file.